\documentclass[pre,floatfix,twocolumn,showpacs,a4paper,nofootinbib]{revtex4}
\usepackage{amsmath}
\usepackage{amsfonts}
\usepackage{amssymb}
\usepackage{calrsfs}
\usepackage{mathrsfs}
\usepackage{epsfig}
\usepackage{graphicx}
\usepackage{dcolumn}
\usepackage{color}

\begin {document}

\title
{Characterization of the community structure in a large-scale production network in Japan}
\author
{
Abhijit Chakraborty, Hazem Krichene, Hiroyasu Inoue, Yoshi Fujiwara  
}
\affiliation
{
\begin {tabular}{c}
Graduate School of Simulation Studies, University of Hyogo, Kobe 650-0047, Japan 
\end{tabular}
}
\begin{abstract}
Inter-firm organizations, which play a driving role in the economy of a country, can be represented in the form of
a customer-supplier network. Such a network exhibits a  heavy-tailed degree distribution, disassortative mixing and 
a prominent community structure. We analyze a large-scale data set of customer-supplier relationships 
containing data from one million Japanese firms. Using a directed network framework,
we show that the production network exhibits the characteristics listed above.
We conduct detailed
investigations to characterize the communities in the network. The topology within smaller communities  
is found to be very close to a tree-like structure but becomes denser as the community size increases. 
A large fraction $(\sim 40\%)$ of firms with relatively small in- or out-degrees have customers or suppliers solely from within their own communities,
indicating interactions of a highly local nature. 
The interaction strengths between communities as measured by the inter-community link weights follow 
a highly heterogeneous distribution. We further present the statistically significant over-expressions
of different prefectures and sectors within different communities.

\end{abstract}

\maketitle

\section {Introduction}

At the mesoscopic level, many real-world networks exhibit community structures,
i.e., structures consisting of groups of nodes that are highly connected 
among themselves and sparsely linked with the other nodes in the network~\cite{girvan2002community,krause2003compartments}.  
These communities are the building blocks of the network and play pivotal roles in its functional activities or dynamic 
processes. For example, in metabolic networks, the communities represent functional groupings of metabolites based on their pathways~\cite{guimera2005functional},
whereas the community structures in air traffic networks represent actual travel patterns~\cite{rosvall2014memory}.
Community structure has diverse applications. It can be used as the basis of an efficient mirror server for achieving
better performance on the World Wide Web~\cite{krishnamurthy2000network}, in the design of powerful recommendation systems~\cite{reddy2002graph} and also to identify
the characteristic features of nodes in subnetworks.  

Over the past decade, extensive studies have been conducted on algorithms for detecting communities~\cite{fortunato2010community, newman2012communities} in which the
number of communities is not required to be fixed a priori, as in graph partitioning problems~\cite{kernighan1970efficient}. 
Many algorithms have been proposed for community detection, and two recent popular techniques are based on modularity optimization~\cite{newman2004finding, newman2004fast}
and the diffusion dynamics in a network~\cite{rosvall2007information,rosvall2008maps}. The modularity is the fraction of intra-community links minus the expected fraction 
given a random distribution. The Newman–Girvan community detection algorithm~\cite{newman2004finding} is a modified divisive clustering technique that 
finds the best partitioning scheme with the optimal modularity using the edge betweenness of the network as a metric.  
This algorithm is very effective in detecting communities, but it is numerically expensive and has limited applicability to small-scale networks. 
Another algorithm~\cite{clauset2004finding}, using an agglomerative hierarchical clustering approach with greedy optimization of the modularity,
has subsequently been proposed for detecting communities in a large-scale network with one million nodes and links. Whereas community detection via modularity 
maximization is based solely on the topology of the network, another popular technique for community detection known as 
``Infomap"~\cite{rosvall2007information,rosvall2008maps} uses the diffusion dynamics in the network. A simple random walk is used to model the diffusion process or
information flow in the network. It is observed that a random walker spends a longer
time within certain subunits of a network. These subunits are the communities of the network. 
Note that the definition of a community in this method is different from that in the modularity maximization method. This technique can capture more complex structures in the network
than can techniques based solely on the network topology. Initially, Infomap was proposed for identifying the two-level partitioning of a network that
minimizes the two-level map equation. Later, the two-level map equation was generalized to the hierarchical map equation, which provides a hierarchical partitioning of a
network~\cite{rosvall2011multilevel}. Whereas modularity-maximization-based techniques are more 
applicable to networks that contain links that represent pairwise relationships, Infomap is more suitable for networks with flows between
nodes~\cite{rosvall2008maps}. 
%---------------------------------------------------------------------------
\begin{figure*}[t]
\begin{center}
\includegraphics[width=0.99\linewidth,clip]{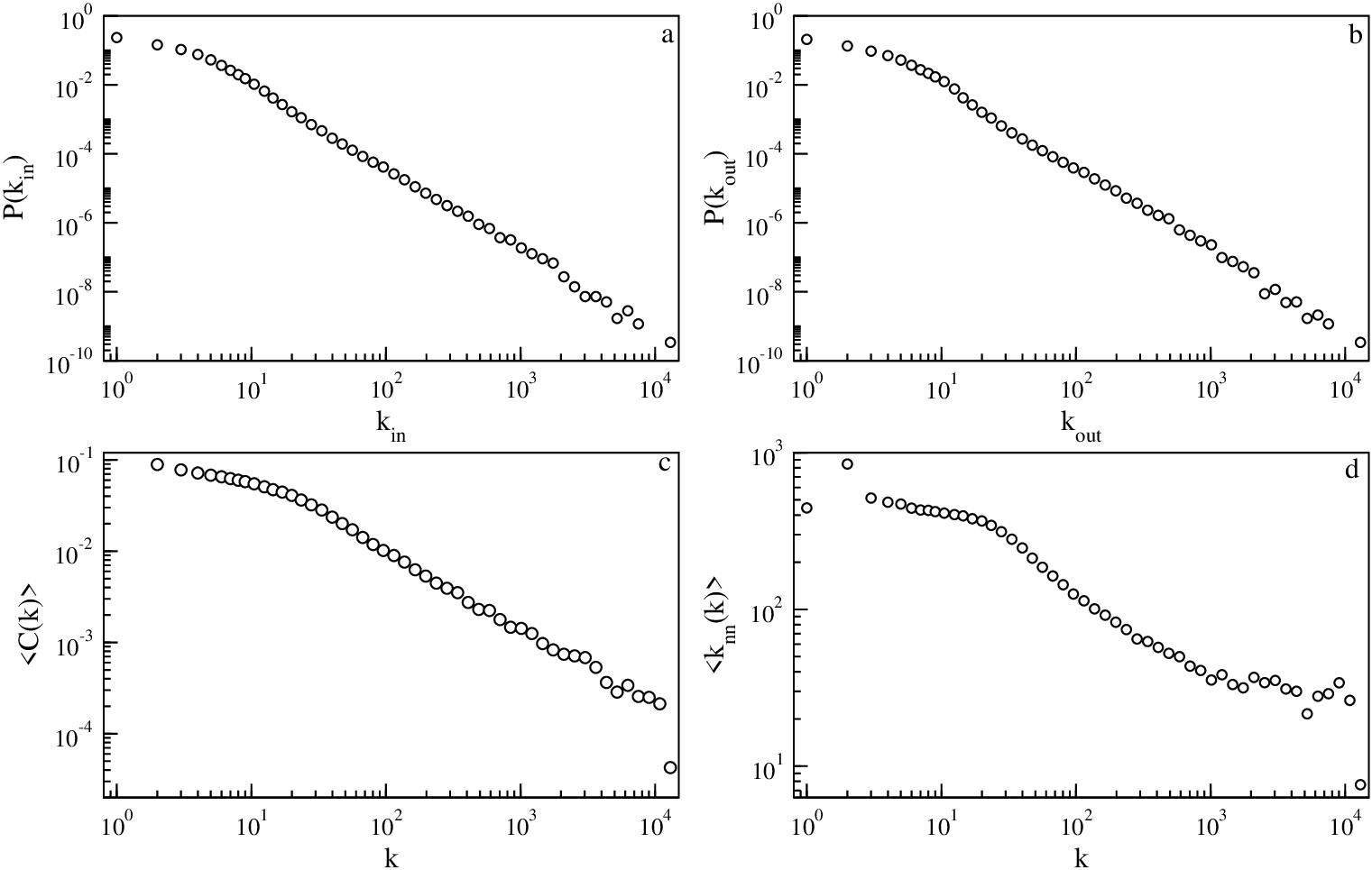}
\end{center}
\caption{(color online)
{\bf Structural properties of the production network.
Probability density distributions $P$ of (a) the nodal in-degrees $k_{in}$ and (b) the nodal out-degrees $k_{out}$.
Variation of (c) the clustering coefficient $C(k) $ as a function of degree $k$
and (d) the average nearest neighbor degree $\langle k_{nn}(k) \rangle$ as a function of degree $k$.}
Logarithmic binning of the horizontal axis is used in all plots. 
%The red line in each plot represents the best power-law fit to the data. 
}
\label{fig1}
\end{figure*}
%---------------------------------------------------------------------------

Although substantial efforts have been devoted to community detection, only a few studies~\cite{lancichinetti2010characterizing, tumminello2011community, tumminello2012identification, vitali2014community, marotta2015bank, marotta2016backbone} 
have characterized the community structure. Here, we characterize the communities in a large-scale production network with one million nodes and links.
Like many social and technological networks, a production network exhibits a prominent community structure~\cite{fujiwara2010large, iino2010community, iino2015community}. 
A production network represents the interactions between firms. It constitutes the backbone of an economy 
in the form of flows of goods and services. Each firm buys intermediate goods from upstream firms for its production, and 
thus, relationships between firms develop through customer-supplier connections.
%A complex network framework is very useful for the analysis of the customer-supplier relationships between firms.
%Firms are the nodes and customer-supplier relationship represents the links of the network. 
The typical topological characteristics of a production network include
a scale-free degree distribution, a small-world nature and disassortative mixing~\cite{fujiwara2010large, mizuno2014structure}. 
The structure of a production network plays an important role in the origin of  business cycles and aggregate fluctuations, as shown by  Acemoglu et al. considering the input-output network of sectors~\cite{acemoglu2012network}. They showed that the asymmetry characterizing the  degree distribution
of the network is the main cause of  large-scale fluctuations in economic systems. Although they found a scale-free nature of 
the degree distribution characterized by a power law in the input-output network of  U.S. industries, such scale-free nature can
disappear when one considers higher levels of coarse-grained input-output networks~\cite{cerina2015world}, which in turn can give an incorrect impression about 
the underlying network topology. In our work, we study the network structure at the firm level, which might be of further importance
in macroeconomic phenomena.
Previous studies on Japanese production networks 
based on modularity maximization have shown that the subcommunities are characterized by geographical regions and sectors~\cite{iino2010community, iino2015community}.
It has also been observed that directed and undirected versions of the same network yield similar results. 
Because the links in a production network represent the flow of goods and services from one firm to another, it is more appropriate to use the Infomap algorithm for community detection.
Moreover, a rigorous, statistically significant characterization of the communities in production networks remains to be conducted.

In this paper, we analyze nationwide inter-firm relationship data from Japan. 
This data set contains data representing one million firms and their customer-supplier relationships. We construct a directed network in which firms are 
represented by nodes and directed links are present from each supplier firm to its customers.
We calculate the standard structural metrics of the network to capture the global properties of the system.
%As the link in this network indicates flow of capital from one firm to another, we use Infomap algorithm to unmask community structure.
By employing the Infomap algorithm, we reveal the community structure of the production network. 
The community structure of the directed network of these firms is found to be distinctly different from that of its undirected counterpart.
The topological features within and among communities indicate a non-trivial local structure of the network. By defining a weighted
directed coarse-grained network in which the nodes represent the identified communities and the numbers of customer-supplier relationships
between communities are treated as link weights, we observe a high heterogeneity of the inter-community interactions. 
Furthermore, in this study, we propose a deeper characterization of the properties of the communities (prefectures and sectors) derived using a 
rigorous statistical procedure, as prescribed in~\cite{tumminello2011community}.

\section {Data}   
\label{data}

Our data consist of $N_o = 1,247,521$ firms and $L_o = 5,488,484$ distinct links representing  customer-supplier relationships between firms throughout Japan 
for the year 2016. The data set is commercially available from Tokyo Shoko Research (TSR), Inc., one of the leading credit research agencies in Japan. 
TSR collects firms' credit information through investigations of financial statements and corporate documents as well as through 
oral surveys at branch offices located across the country. The data set contains the precise geographic locations and sectorwise classifications of the 
firms. 

We use a prefecture-based division of all regions to analyze our data. All $47$ prefectures 
are listed in Appendix A. 
To see whether the sectors to which different firms belong play any role in community formation, we use the Japan Standard Industrial Classification\footnote{\url{http://www.soumu.go.jp/toukei_toukatsu/index/seido/sangyo/}}, which
divides all the firms into 98 major groups. All 98 major sectors 
are listed in Appendix B.

%---------------------------------------------------------------------------
\begin{figure}[t]
\begin{center}
\includegraphics[width=0.99\linewidth,clip]{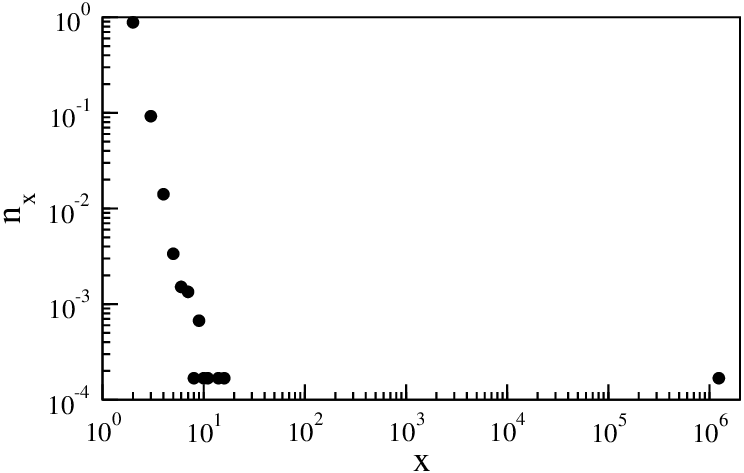}
\end{center}
\caption{
{\bf Distribution $n_x$ of the component sizes $x$ in the network.
The largest weakly connected component contains $\sim 99\%$ of the nodes in the entire network.}
}
\label{fig2}
\end{figure}
%---------------------------------------------------------------------------  

\section{Results}
We reveal the detailed topological features of the Japanese production network. To define the network,
each firm is treated as a node, and a directed link of the form $A \rightarrow B$ indicates that $A$ is a supplier firm
for firm $B$. The structure of the production network exhibits a number of empirical patterns, which we present in the following
subsections. 

\subsection{Structural properties of the production network}
The links in the production network are directed in nature, which allows us to characterize the nodes in terms of their in- and out-degrees.
The in-degree and out-degree of a node represent its numbers of suppliers and customers, respectively. We observe that both
the in-degree and out-degree distributions have a heavy-tailed nature, as shown in Fig.~\ref{fig1}~(a-b). 
%the best fits of these distributions to a power-law decay have a functional form of $P(k_{in/out})\sim k^{-\gamma_{in/out}}$, with $\gamma_{in} = 2.37$ and $\gamma_{out} = 2.21$. 
It is also observed that the in- and out-degrees of the nodes are positively
correlated. 
%Our results reflect the scale-free nature of the production network, with 
The heavy-tailed behavior of the degree distribution reflects the fact that production network has
a large number of firms with few suppliers
or customers and a smaller number of firms with many suppliers or customers.
Both the in-degree and out-degree show a positive and significant correlation with the firm size measured by using either its net 
sales or total numbers of employees~\cite{fujiwara2010large}.  Because large firms  generally have more suppliers and customers 
than small firms, the rich-get-richer principle is the key to such a power law nature of the degree distributions~\cite{barabasi1999emergence, chakraborty2010weighted}. Because of the high asymmetry in the degree 
distributions of the underlying network, aggregate fluctuations can appear in the system due to idiosyncratic shocks to large firms~\cite{acemoglu2012network}. 

We further investigate the cliqueness among neighbors and the mixing properties of the network considering undirected links. 
The cliqueness among the neighbors of a node is measured by the clustering coefficient, which is also a measure of the three-point
correlations among neighboring nodes. As seen from Fig.~\ref{fig1}~(c), the clustering coefficients $C(k)$ for 
the production network decay with $k$ 
%following a power law, with the form $C(k) \simeq k^{-\beta}$ with $\beta = 0.88$, 
implying the presence of a hierarchical structure in the network. The mixing patterns of the nodes are measured by the average degree $\langle k_{nn}(k) \rangle$
of an arbitrary neighbor of a node of degree $k$. Fig.~\ref{fig1}~(d) shows that $\langle k_{nn}(k) \rangle$ decreases 
%following a power law 
as $k$ increases,
%with the form  $\langle k_{nn}(k) \rangle \simeq k^{-\nu}$ with $ \nu = 0.58$, 
indicating a disassortative nature of the
network, i.e., nodes of higher degrees are connected to nodes of lower degrees. This finding reflects the fact that large firms are generally
connected with small firms. 

The component size distribution is another important property of the network. The component size is defined as the number of nodes 
in a subnetwork in which at least one path exists between any arbitrary pair of nodes.
It is observed that $99\%$ of the nodes are contained within the largest weakly connected component of the network, whereas the sizes of the other components are very small, as shown in Fig.~\ref{fig2}. 
More precisely, the largest weakly connected component contains  $ N = 1, 234, 687$ nodes and  $ L = 5, 481, 403$ links.

%%---------------------------------------------------------------------------
%\begin{figure}[top]
%\begin{center}
%\includegraphics[width=0.99\linewidth,clip]{Pk_meta.eps}
%\end{center}
%\caption{(color online)
%{\bf Degree distributions $P(k)$ and strength distributions $P(s)$ of coarse the grained network.}
%}
%\label{fig4}
%\end{figure}
%---------------------------------------------------------------------------
%---------------------------------------------------------------------------
\begin{figure}[t]
\begin{center}
\includegraphics[width=0.99\linewidth,clip]{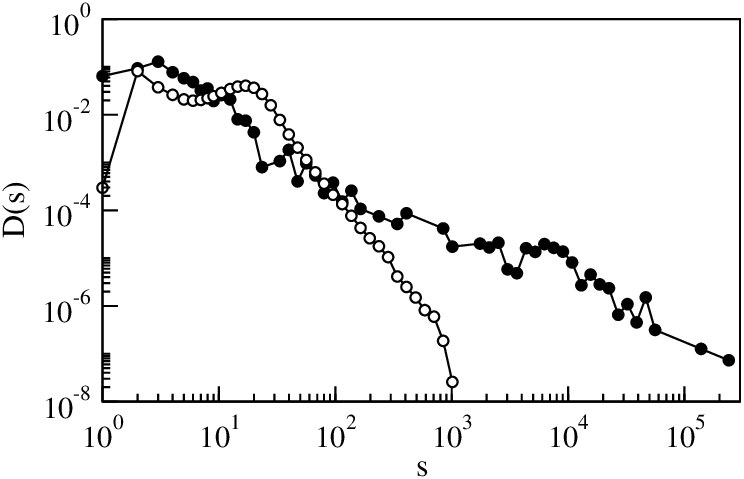}
\end{center}
\caption{
{\bf Distributions $D(s)$ of community sizes $s$ for the empirical production network with directed links (filled circles) and for its randomized counterpart (open circles).
%(b) rank-size distribution of communities for undirected network of firms.
} 
Communities are detected in the largest weakly connected component of the network.
Logarithmic binning is used for the horizontal axis. 
}
\label{fig3}
\end{figure}
%---------------------------------------------------------------------------  
\subsection{Community structure of the production network}
Although the structural properties discussed above are important for obtaining a general understanding of the system, a complex network such as this one also exhibits a far more
complex pattern, namely, a community structure. We perform a community detection analysis on the largest weakly connected component of the production network, 
considering directed links. We use the hierarchical map equation~\cite{rosvall2011multilevel} to discover the community structure because this method is very efficient for
large-scale complex networks and is also applicable to directed networks. In this procedure, a random walker is used as a proxy for 
the flow of goods and services between firms in the production network, and the algorithm identifies the best hierarchical partitioning with the shortest 
average per-step description length (code length) for the random walker. 

The analysis of the empirical production network reveals $311$ communities and 
$8054$ inter-community links at the top community level with a code length of $12.925$. To determine the statistical significance of our result, we compare it with the result obtained from an identical 
analysis of a `null model', which is a random network with the same nodal degree values $\{k_{in}(i)\}$ and $\{k_{out}(i)\}$.
Following the the method of configuration models~\cite{newman2010networks}, we construct the randomization version of the empirical network using a pair-wise link exchange technique with 
the constraint of no multiple links between any pairs of nodes. 
In stark contrast with the empirical result, the randomized version of the empirical network is found to contain $ 62, 917 \pm 120$ communities and $4, 234, 729 \pm 1, 552$ inter-community connections with
a code length of $14.214 \pm 0.005$. The results are averaged over 10 uncorrelated random networks. 
As seen from Fig.~\ref{fig3}, the community size distributions for the empirical network and the randomized version of the empirical 
network are found to be distinctly different in nature. Whereas the empirical network has a broad distribution of community sizes with a wide range of values 
spanning several orders of magnitude, the randomized version of the empirical network has a comparatively narrow distribution, in which the largest community
size is $\sim 1, 000$. 
We further compare the community structure of the directed network of firms with that of its undirected counterpart. In the undirected network, we find
only $26$ communities with distinct and widely varying sizes (not shown), which is smaller than the number
of communities in the directed network. This difference in the community structure findings between the directed and undirected versions of the production network 
arises because Infomap allows the random walker to move only unidirectionally on directed links and permits two-way movement on undirected links.
A random walker on an undirected network can move from one sub-region  of the network to another sub-region  with a transition probability 
proportional to the total number of connected links between the two sub-regions. For the same movement on the directed version of the network, 
the transition probability is proportional to the total number of outgoing links from one region  to another.  Clearly, the 
transition probability for the movement between sub-regions is always higher on an undirected network than on its directed counterpart. 
For this reason, in many cases, distinctly different communities in directed networks can be merged with the same community in 
an undirected network. As a result, we find more communities of smaller size on directed networks than on the 
undirected version of the network. 
This is not a very special case with the production network.
The only distinct property of our network is the low fraction ($2.68 \%$) of the bi-directional links. 
The directed version of the production network shows different community structure from its undirected  counterpart because of a low fraction of bi-directional links.

%---------------------------------------------------------------------------
\begin{figure}[t]
\begin{center}
\includegraphics[width=0.99\linewidth,clip]{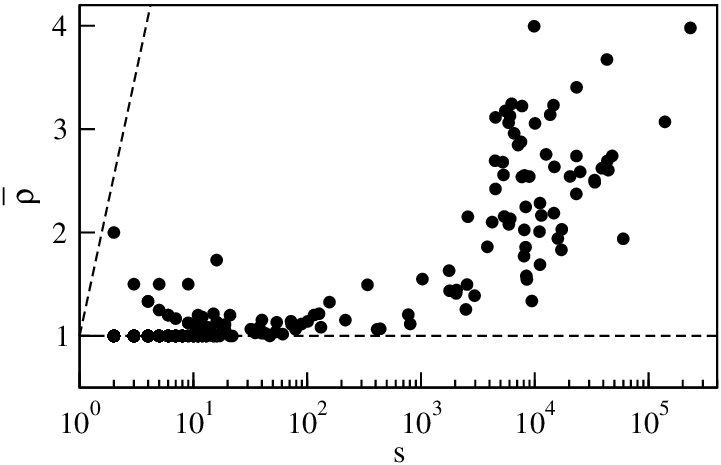}
\end{center}
\caption{
%(color online)
{\bf Scatter plot of the intra-community scaled link density $\bar{\rho}$ versus the community size $s$.}
The dotted lines represent two limiting cases: $\bar{\rho} = s$ (clique) and $\bar{\rho} = 1$ (tree).
}
\label{fig4}
\end{figure}
%---------------------------------------------------------------------------

\subsubsection{Topological features of communities in the production network}
We investigate the topological features within each community in the production network. The link density within a community is
the ratio of the number of internal links to the maximum possible number of links. The link density $\rho$ within a community of
size $s$ can be calculated as  $\rho = e/s(s-1)$, where $e$ is the number of links within the community. The scaled link density $\bar{\rho}$
within a community is defined as $\bar{\rho} = \rho s = e/(s-1)$. A value of $\bar{\rho} = 1$ corresponds to a community with a tree-like structure
with unidirectional links, and $\bar{\rho} = s$ corresponds to a complete graph structure, i.e., a structure in which every node is connected to all other nodes in the community. 
Fig.~\ref{fig4} shows a scatter plot of the scaled link density $\bar{\rho}$ versus the community size $s$. It is evident that the network structures within the
communities are far from being complete graph structures; indeed, they are very close to an ideal tree-like structure when the community size is small $(s < 1000)$.
However, beyond $(s > 1000)$, the scaled link density gradually increases as the community size increases. 

\begin{figure}[t]
\begin{center}
\includegraphics[width=0.99\linewidth,clip]{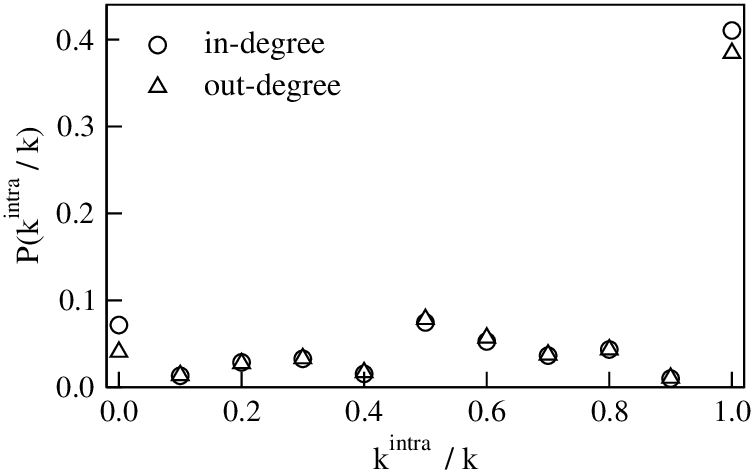}
\end{center}
\caption{
%(color online)
{\bf Probability distributions $P(k_{in}^{intra} / k_{in})$ and  $P(k_{out}^{intra} / k_{out})$ for the fractions of in-neighbors ($k_{in}^{intra} / k_{in}$) and out-neighbors ($k_{out}^{intra} / k_{out}$) 
of a node that belong to the same community as that node.}
The circle and triangle symbols represent the in- and out-neighbor distributions, respectively. 
}
\label{fig5}
\end{figure}

Next, we study the fraction of the neighbors of a node that belong to its own community. We find this fraction separately for the in-degree ($k_{in}^{intra} / k_{in}$) 
and out-degree ($k_{out}^{intra} / k_{out}$)  of the node. 
%In and out degrees of a node here indicate suppliers and customers of a firm respectively.
The probability distributions for the fraction of suppliers ($P(k_{in}^{intra} / k_{in})$) and the fraction of customers ($P(k_{out}^{intra} / k_{out})$) of a node
that belong to its community are shown in Fig.~\ref{fig5}. The two distributions are very similar in nature and have a maximum value of $~0.4$ at
$k_{in}^{intra} / k_{in} = k_{out}^{intra} / k_{out}=1 $, meaning that for $40\%$ of the firms, all suppliers and customers of a firm belong to own community. 
We further observe that these firms have very small ($<100$) in- or out-degrees.  
This result reflects the fact that for a large proportion of firms, their interactions are confined within a local region.

%---------------------------------------------------------------------------
%---------------------------------------------------------------------------
%---------------------------------------------------------------------------
\begin{figure}[t]
\begin{center}
\includegraphics[width=0.99\linewidth,clip]{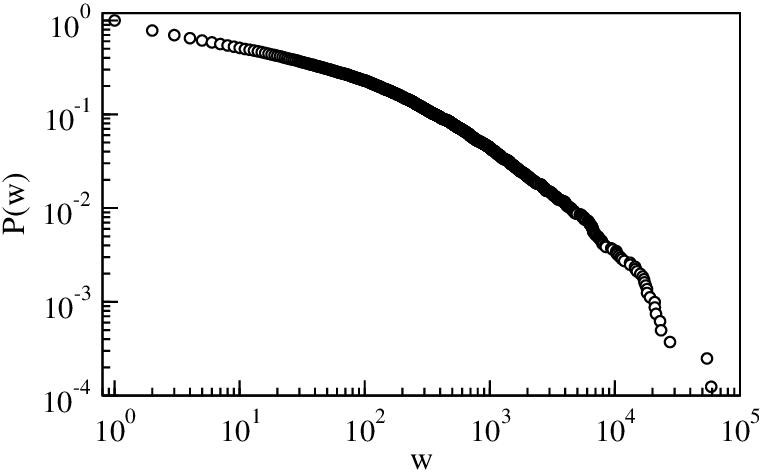}
\end{center}
\caption{(color online)
{\bf The complementary cumulative distribution function $P(w)$ of the inter-community link weights $w$ for the production network.}
A power-law fit to the data (red line) using the maximum likelihood estimation technique yields $P(w)\sim w^{-\gamma_w +1}$ 
with $\gamma_w =2.00 \pm 0.10$, $w_{min}=642 \pm 301$, and $p~value = 0.07$.
}
\label{fig6}
\end{figure}
%--------------------------------------------------------------------------- 

%------------------------------------------------------------------------------------------------------
\begin{table*}[t] 
\begin {center}
 \scriptsize
\begin{tabular}{|l|l|p{7.5cm}|p{7.5cm}|} \hline
  Rank   & Size      & Over-expression of sectors            &  Over-expression of prefectures  \\ \hline
  1      & 233, 294   & Manufacturing, Electronics, Water transport, Wholesale, etc.   &  Urban prefectures (Tokyo and its neighboring prefectures, Osaka, Aichi, Hyogo, etc.)          \\ 
  2      & 139, 380   & Agriculture, Food, Fisheries, Road freight transport, Cooperative associations, N.E.C., etc.        & Rural prefectures (Aomori, Miyagi, Shizuoka, etc.)             \\
  3      &  59, 906   & Construction, Real estate, Banking, etc.      &  Tokyo and its neighboring prefectures, Osaka          \\
  4      &  47, 849   & Manufacture of textiles, rubber, leather, etc.        &  Tokyo, Osaka, Kyoto, Aichi, etc.           \\ 
  5      &  44, 349   & Medical services, Research institutes, Chemical products, etc.     &  Hokkaido, Tokyo, Hiroshima, etc.     \\                  
  6      &  43, 397   & Retail trade (machinery and equipment), Automobile  maintenance, Transport, Insurance institutions, etc.       & Many (22) prefectures            \\                
  7      &  43, 018   & Multiple sectors        & Hokkaido              \\         
  8      &  38, 819   & Multiple sectors       & Tokyo               \\ 
  9      &  33, 654   & Information services and many others   & Tokyo, Kanagawa, Osaka            \\
  10     &  33, 563   & Construction and others         & Gifu, Aichi, Mie                         \\ \hline
     \end{tabular}
\caption{Brief summary of our results on the over-expression of sectors and prefectures in the ten largest communities}
\label{table1}
\end {center}
\end {table*}
%---------------------------------------------------------------------------------
%---------------------------------------------------------------------------
%\begin{figure}[top]
%\begin{center}
%\includegraphics[width=0.99\linewidth,clip]{Dominant_sec_geo.eps}
%\end{center}
%\caption{
%{\bf Scatter plot between dominant prefecture share and dominant sector share in the communities of the network.}
%Symbol sizes are proportional to the logarithmic of community sizes. 
%}
%\label{fig6}
%\end{figure}
%--------------------------------------------------------------------------- 

%Fig.~\ref{fig5} shows the dominance of geography and sector share over all communities having size more than 10. 
%The size of the circle indicates size of the communities in logarithmic scale. 
%The position of the most of the communities below the diagonal implying a clear geography dominance over the sector share dominance. 
%---------------------------------------------------------------------------
%\begin{figure}[top]
%\begin{center}
%\includegraphics[width=0.99\linewidth,clip]{Demply.eps}
%\end{center}
%\caption{(color online)
%{\bf Distributions $D(e)$ of number of employees e for a firm  in the network (black circle) and for communities in the coarse grained network (red circle).}
%}
%\label{fig7}
%\end{figure}

% The coarse grained version of the production network, in which one considers communities as nodes, is very sparse in nature as the 
%number of inter-modular links and communities are of the same order.

An inter-community link indicates that at least one customer-supplier relationship exists between
the members of the two linked communities. These links are directed, and the corresponding in- and out-degree distributions exhibit very different behaviors compared with the original
network. We also investigate the link weights of the inter-community links, where a link weight is defined based on the total number of customer-supplier relationships between
members of the two linked communities. Fig.~\ref{fig6} shows the complementary cumulative distribution function $P(w)$ of the inter-community link weights $w$, which is heavy tailed in nature.
However, A power-law fit to the data using the maximum likelihood estimation technique~\cite{clauset2009power} yields $P(w)\sim w^{-\gamma_w +1}$ 
with $\gamma_w =2.00 \pm 0.10$, $w_{min}=642 \pm 301$, and $p~value = 0.07$. The low $p~value$ implies that the distribution is not a power law.   
%a power-law decay of the functional form $P(w)\sim w^{-\gamma_w+1}$ with $\gamma_w =2.00 \pm 0.10$ 
%is obtained for this distribution via maximum likelihood estimation~\cite{clauset2009power}. 
The distribution of the inter-community link weights indicates highly heterogeneous relationships between the communities in the production network.

\subsubsection{Over-expression of node attributes in the production network}

Most real-world networks are characterized by a community structure.  However, the factors that influence 
such clustering of nodes are different for different systems. For example,  the community
structure of a social network for mobile phone communications shows that people speaking the same 
language belong to the same community~\cite{blondel2008fast},  biological functions play a key role in forming 
communities of proteins in yeast~\cite{chen2006detecting}, and in stock markets, communities present similarities in the 
economic sectors of the stocks~\cite{onnela2003dynamics}. The communities in a network are mainly formed following
the principle of homophily. We use the over-expression analysis to characterize the communities. 
The study helps one to understand the basis of forming communities that have particular attributes.
We consider geographical and sector attributes; however, in general, one can take any attributes.

%---------------------------------------------------------------------------
\begin{figure}[t]
\begin{center}
\includegraphics[width=0.99\linewidth,clip]{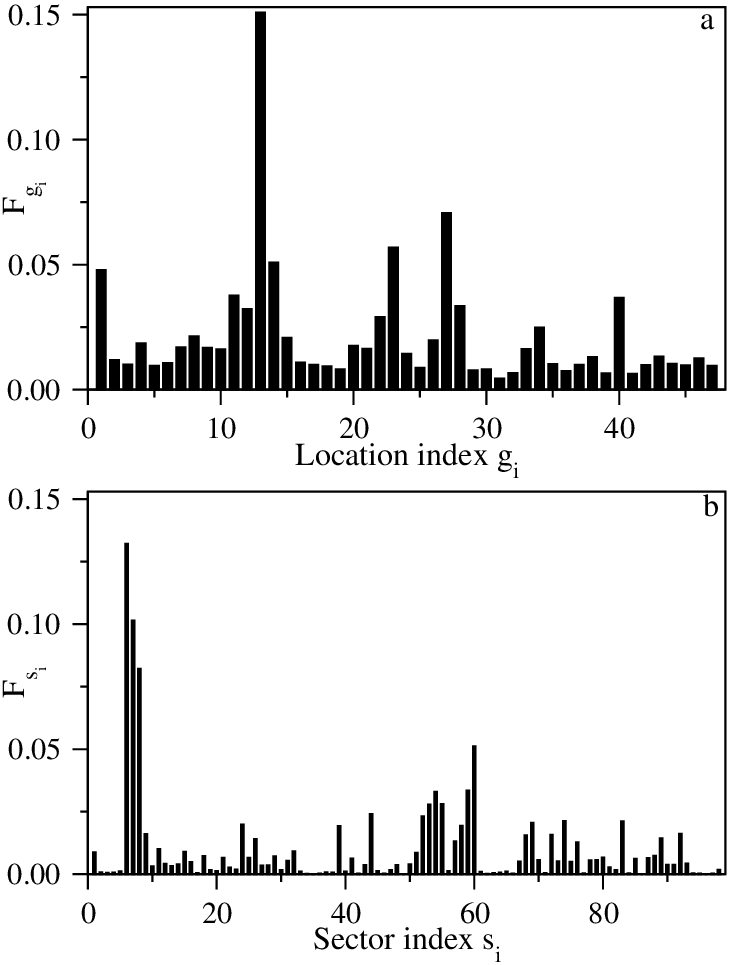}
\end{center}
\caption{
{\bf The location and sectorial attributes of firms.
(a) The fraction of firms belonging to each prefecture, $F_{g_i}$, is plotted versus the corresponding index $g_i$.
(b) The fraction of firms belonging to each sector, $F_{s_i}$, is plotted versus the corresponding index $s_i$.}
The location indices $g_i$ and the sector indices $s_i$ are defined in Appendix A and Appendix B, respectively. 
}
\label{fig7}
\end{figure}
%--------------------------------------------------------------------------- 

%---------------------------------------------------------------------------
\begin{figure}[t]
\begin{center}
\includegraphics[width=0.99\linewidth,clip]{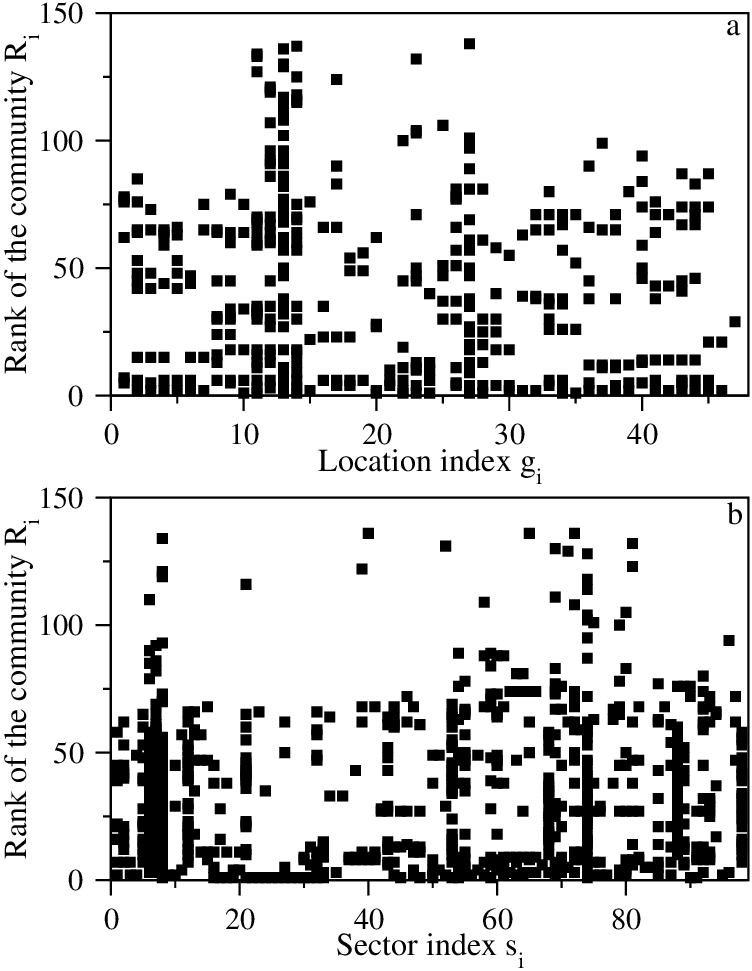}
\end{center}
\caption{
{\bf Statistically significant over-expression of (a) prefectures and (b) sectors in individual communities, represented by black square symbols.}
The vertical axis represents the size-wise rank $R_i$ of the $i$-th community.}
\label{fig8}
\end{figure}
%---------------------------------------------------------------------------
We investigate the over-expression of different prefectures or sectors within a community following the procedure used in~\cite{tumminello2011community}.
The probability that $X$ randomly selected elements from a cluster $C$ of size $N_C$ will have attribute $Q$ is measured by the hypergeometric distribution
$$H(X|N, N_C, N_Q)= \frac{\binom{N_C}{X}\binom{N-N_C}{N_Q-X}}{\binom{N}{N_Q}}$$,
where $N_Q$ represents the total number of elements with attribute $Q$ in the system.
If $N_{C,Q}$ is the number of elements in cluster $C$ with attribute $Q$, then one can calculate an associated 
{\em p value:} $$p(N_{C,Q}) = 1- \sum_{X=0}^{N_{C,Q}-1}H(X|N, N_C, N_Q)$$ 
The attribute $Q$ is over-expressed within a community if $p(N_{C,Q})$ is smaller than some threshold value $p_c$.
We must choose $p_c$ appropriately to exclude false positives since we are considering a multiple-hypothesis test.
We set $p_c=0.01/N_A$, as used in~\cite{tumminello2011community}, which is sufficient for Bonferroni correction~\cite{miller1981normal}.  

In our study, $N_A$ denotes the total number of possible attributes and takes on values of $47$ and $98$ for the prefecture- and sector-based attributes, respectively.
Fig.~\ref{fig7}~(a) shows the fraction of firms belonging to each prefecture, $F_{g_i}$. The largest proportion of the firms $(15.0\%)$
belong to Tokyo, the capital of Japan.  
Fig.~\ref{fig7}~(b) shows the fraction of firms belonging to each sector, $F_{s_i}$. 
The largest proportion of the firms $(13.2\%)$ belong to the general construction work sector, which includes both public and private construction work.   
The over-expression of prefectures and sectors within individual communities consisting of at least 10 firms is visualized in Fig.~\ref{fig8}. Prefectures are over-expressed within
$138$ communities in the production network, and sectors are over-expressed within $136$ communities. A brief summary of the over-expression of prefectures and sectors in the ten largest communities is tabulated
in Table~\ref{table1}, and the details are given in Appendix C. Our investigation reveals that the largest community is characterized by the over-expression of manufacturing sectors and urban prefectures.
The agriculture, food, and fisheries sectors and rural prefectures are the defining characteristics of the second-largest community. The third-largest community is characterized by the construction, real estate, and banking sectors and by 
Osaka and Tokyo and its neighboring prefectures. The fourth-largest community is mainly characterized by the manufacturing sectors of textiles, rubber, and leather. Medical
and other health services constitute the most distinctive feature of the fifth-largest community. Retail trade (machinery and equipment) and automobile maintenance services dominate
 the sixth-largest community. Hokkaido is the only over-expressed prefecture in the seventh-largest community, whereas the eighth-largest community is strongly associated with Tokyo.
Only three prefectures --- Tokyo, Kanagawa, and Osaka --- are over-expressed in the ninth-largest community, and three different prefectures --- Gifu, Aichi, and Mie --- are the distinctive
features of the tenth-largest community. It is observed that for a given community, its over-expressed sectors are strongly related to its over-expressed prefectures.
We also note that even if the extent to which an attribute is present in the system as a whole is very small, it may be over-expressed in a particular community. For example, the banking
sector is represented by only $161$ banks in our data set, but this sector is over-expressed in the third- and ninth-largest communities. We conclude that different communities are characterized
by distinct features related to different sectors and prefectures.

%we investigate the firm size distribution with the number of employees as the measure of firm size. We have only $7\%$ missing data in our data set.
%We have studied the firm size distribution for both the original network and coarse grained network. As can be seen from Fig.~\ref{fig6} distributions 
%of firm size follow the well-known Zipf's law $D(e)\sim e^{-\alpha}$ for both the networks. The slope of the straight portion of the distributions in double
%log scale gives the value of $\alpha = 2.09 \pm 0.02$ and $2.0 \pm 0.06 $ respectively for the original network and its coarse grained version.
%Note that the tail of the distribution in the coarse grained network is shifted up from its intermediate straight portion because of finite size effect.   
%From this above result, we conclude that characteristic of the firm size distributions remain invariant even in the coarse grained version of the network 
%and follows Zipf's law. 
 
\section {Conclusion} 
We have studied a large-scale production network in Japan that exhibits 
a  heavy-tailed degree distribution, hierarchical clustering and disassortative
mixing, consistent with previous studies~\cite{fujiwara2010large, mizuno2014structure}.
We have delved deeper into the data using the Infomap technique to detect the 
communities within the network. This production network
contains many communities of highly heterogeneous sizes and with a wide range of
values of the link weights between them. The community structures identified in the directed and undirected
versions of the network display distinct behaviors. Smaller communities tend to have tree-like structures, whereas larger communities have higher link densities.
A large fraction ($40\%$) of the firms in the production network have suppliers or customers
only from within their own communities. We have also characterized the communities based on the over-expression of geographical locations and sectorial classifications. 

The topological properties of a network play a crucial role in dynamical processes.
Our analysis presents the backbone that can be utilized to study any
agent based modeling to explain economic phenomena. Our analysis 
further shows sectorial and regional attributes of firms have an 
important role in forming communities.  This indicates the network 
structure of firms is neither random nor because of the preferential 
attachment rule~\cite{barabasi1999emergence}. 
A future research in this direction will be to model the production network that shows community structure with homophily in industrial sectors and regions.
This study has an application to understand the business cycle correlations in production network~\cite{krichene2017business}. 

%It is found that 
%geographical dominance is more visible than sectorial influence. It is also
%observed that the firm size distribution in the production network as well as 
%in its coarse grained version follows Zipf's law robustly. 

Our results suggest several interesting topics for future economic research. 
In this study, we have characterized the communities only at the top community level.
One could generalize the applied technique to characterize the communities at all levels.
%The highly heterogeneous  community's number reflects the 
A bottom-up structural analysis of the Japanese economy would be helpful for better understanding its complexity. 
Thus, future economic research seeking to understand GDP
fluctuations, inflation and monetary stability should evaluate the Japanese economy at the microscopic
level. 
%Moreover, the Zipf's law of firm size distribution
%indicates the heterogeneous impact of firms on the Japanese economy. 
Accordingly,
future studies should investigate the roles and risks of the largest firms
to yield an understanding of economic shock and driven systemic risk.

\noindent
\begin{center}
{\bf ACKNOWLEDGMENTS}
\end{center}
This research was supported by MEXT as Exploratory Challenges on
Post-K computer (Studies of Multi-level Spatiotemporal Simulation of
Socioeconomic Phenomena), and the Grant-in-Aid for Scientific Research (KAKENHI) by JSPS Grant Number 17H02041.
We thank Hideaki Aoyama, Hiroshi Iyetomi and Yuichi Kichikawa for useful discussions on Infomap. 

%\noindent
\begin{center}
{\bf APPENDIX A: GEOGRAPHICAL ATTRIBUTES OF FIRMS}
\end{center}
Here we list the $47$ prefectures in Japan with their index, that we have used for our study:

1. Hokkaido, 2. Aomori, 3. Iwate, 4. Miyagi, 5. Akita, 6. Yamagata, 7. Fukushima, 8. Ibaraki, 9. Tochigi, 10. Gunma,
11. Saitama, 12. Chiba, 13. Tokyo, 14. Kanagawa, 15. Niigata, 16. Toyama, 17. Ishikawa, 18. Fukui, 19. Yamanashi, 20. Nagano,
21. Gifu, 22. Shizuoka, 23. Aichi, 24. Mie, 25. Shiga, 26. Kyoto, 27. Osaka, 28. Hyogo, 29. Nara, 30. Wakayama,
31. Tottori, 32. Shimane, 33. Okayama, 34. Hiroshima, 35. Yamaguchi, 36. Tokushima, 37. Kagawa, 38. Ehime, 39. Kochi, 40. Fukuoka,
41. Saga, 42. Nagasaki, 43. Kumamoto, 44. Oita, 45. Miyazaki, 46. Kagoshima, 47. Okinawa.

\begin{center}
{\bf APPENDIX B: SECTORIAL ATTRIBUTES OF FIRMS}
\end{center}
Here we list  the $98$ major sectors in Japan with their index using Japan Standard Industrial Classification:

1. Agriculture, 2. Forestry, 3. Fisheries, except Aquaculture, 4. Aquaculture, 5. Mining and quarrying of stone, 6. Construction work, general including public and private construction work,
7. Construction work by specialist contractor, except equipment installation work, 8. Equipment installation work, 9. Manufacture of food, 10. Manufacture of beverages, tobacco and feed,
11. Manufacture of textile products 12. Manufacture of lumber and wood products, except furniture, 13. Manufacture of furniture and fixtures
14. Manufacture of pulp, paper and paper products, 15. Printing and allied industries, 16. Manufacture of chemical and allied product, 17. Manufacture of petroleum and coal products,
18. Manufacture of plastic products, except otherwise classified, 19. Manufacture of rubber products, 20. Manufacture of leather tanning, leather products and fur skins,
21. Manufacture of ceramic, stone and clay products, 22. Manufacture of iron and steel, 23. Manufacture of non-ferrous metals and products, 24. Manufacture of fabricated metal products,
25. Manufacture of general-purpose machinery, 26. Manufacture of production machinery, 27. Manufacture of business oriented machinery, 28. Electronic parts, devices and electronic circuits,
29. Manufacture of electrical machinery, equipment and supplies, 30. Manufacture of information and communication electronics equipment, 31. Manufacture of transportation equipment,
32. Miscellaneous manufacturing industries, 33. Production, transmission and distribution of electricity, 34. Production and distribution of gas, 35. Heat supply,
36. Collection, purification and distribution of water and sewage collection, processing and disposal, 37. Communications, 38. Broadcasting, 39. Information services,
40. Services incidental to internet, 41. Video picture information, sound information, character information production and distribution, 42. Railway transport, 43. Road Passenger transport,
44. Road freight transport, 45. Water transport,  46. Air transport, 47. Warehousing, 48. Services incidental to transport, 49. Postal services, including mail delivery, 
50. Wholesale trade, general merchandise, 51. Wholesale trade (textile and apparel), 52. Wholesale trade (food and beverages), 53. Wholesale trade (building materials, minerals and metals, etc),
54. Wholesale trade (machinery and equipment), 55. Miscellaneous Wholesale trade, 56. Retail trade, general merchandise, 57. Retail trade (woven fabrics, apparel, apparel accessories and notions),
58. Retail trade (food and beverage), 59. Retail trade (machinery and equipment), 60. Miscellaneous retail trade, 61. Nonstore retailers, 62. Banking, 63. Financial institutions for cooperative organizations,
64. Non-deposit money corporations, including lending and credit card business, 65. Financial products transaction dealers and futures commodity dealers, 66. Financial auxiliaries,
67. Insurance institutions, including insurance agents brokers and services, 68. Real estate agencies, 69. Real estate lessors and managers, 70. Goods rental and leasing,
71. Scientific and development research institutes, 72. Professional services, N.E.C., 73. Advertising, 74. Technical services, N.E.C., 75. Accommodations, 
76. Eating and drinking places, 77. Food take out and delivery services, 78. Laundry, beauty, and bath services, 79. Miscellaneous living-related and personal services, 
80. Services for amusement and recreation, 81. School education, 82. Miscellaneous education, learning support, 83. Medical and other health services, 84. Public health and hygiene,
85. Social insurance, social welfare and care services, 86. Postal services, 87. Cooperative associations, N.E.C, 88. Waste disposal business, 89. Automobile maintenance services,
90. Machine, etc. repair services, except otherwise classified, 91. Employment and worker dispatching services, 92. Miscellaneous business services, 93. Political, business and cultural organizations, 
94. Religion, 95. Miscellaneous services, 96. Foreign governments and international agencies in Japan, 97. National government services, 98. Local government services. 

\begin{center}
{\bf APPENDIX C: DETAIL DESCRIPTIONS OF TEN LARGEST COMMUNITIES}
\end{center}
Here we report the details of over-expression of prefectures and sectors in ten largest communities.
The pair of values within the parentheses indicate the number of occurrence of the attribute within the community 
and in the whole system. 
%--------------------------------------------------------------------------- 
\begin {itemize}
\item {\bf Community rank:} 1, {\bf size:} 233294\\
{\bf Prefecture over-expressions:} Gunma (4436/19893), Saitama (11181/46418), Tokyo (37172/186179), 
Kanagawa (15776/62781), Nagano (4528/21641), Shizuoka(7889/35806), Aichi (19903/70128)
Mie (3771/17747), Osaka ( 26641/87226), Hyogo (10120/41282), Hiroshima (6795/30651), Yamaguchi (2881/12604).\\
{\bf Sector over-expressions:} Equipment installation work(24461/101430), Manufacture of chemical and allied product (2420/5946),
Manufacture of petroleum and coal products (198/518), Manufacture of plastic products except otherwise classified (6030/8922),
Manufacture of rubber products (1530/2060), Manufacture of ceramic, stone and clay products (1846/8051), Manufacture of iron and steel (2602/3210),
Manufacture of non-ferrous metals and products (1960/2291), Manufacture of fabricated metal products (17655/24494),
Manufacture of general-purpose machinery (6676/8125), Manufacture of production machinery (13814/17342),
Manufacture of business oriented machinery (2672/4194), Electronic parts, devices and electronic circuits (3768/4363)
Manufacture of electrical machinery, equipment and supplies (7081/8795), Manufacture of information and communication electronics equipment (1389/2101)
Manufacture of transportation equipment (5049/6623), Miscellaneous manufacturing industries (2585/11270),
Production, transmission and distribution of electricity (376/1306), Water transport (894/1490), Services incidental to transport (1219/4474),
Wholesale trade, general merchandise (1568/4891), Wholesale trade (building materials, minerals and metals, etc) (10858/34380), 
Wholesale trade (machinery and equipment) (19336/40672), Retail trade (machinery and equipment) (11653/41319),
Technical services, N.E.C. (5298/26216), Machine, etc. repair services, except otherwise classified (2490/4670)
Employment and worker dispatching services (1055/4683).
\item {\bf Community rank:} 2, {\bf size:} 139380\\
{\bf Prefecture over-expressions:} Aomori (2447/14681), Iwate(2115/12419), Miyagi (3388/22888), Akita (1804/11705),
Yamagata (1969/13168), Fukushima (2596/20897), Niigata (3153/25586), Nagano (2914/21641), Shizuoka (4713/35806), Wakayama (1341/10004)
Tottori (920/5468), Shimane (1030/8209), Tokushima (1268/9172), Kagawa (2018/12287), Ehime (2271/16097), Kochi (1125/8078),
Saga (1468/7873), Nagasaki (1945/12078), Kumamoto (2356/16358), Oita (2129/12738), Miyazaki (2153/12013), Kagoshima (3251/15423).\\
{\bf Sector over-expressions:} Agriculture (4277/10747), Fisheries, except aquaculture (563/724), Aquaculture (627/771),
Manufacture of food (16020/19833), Manufacture of beverages, tobacco and feed (2675/3935), Road freight transport (3615/29678),
Warehousing (372/1997), Wholesale trade, general merchandise (931/4891), Wholesale trade(food and beverages) (23126/28535),
Miscellaneous wholesale trade (6497/34637), Retail trade, general merchandise (1095/1531), Retail trade (food and beverage) (18082/23948),
Nonstore retailers (297/1203), Financial institutions for cooperative organizations (130/511), Non-deposit money corporations, 
including lending and credit card business (180/755), Real estate lessors and managers (3631/25301), Accommodations (2339/6062),
Eating and drinking places (11276/15725), Food take out and delivery services (280/432), Miscellaneous living-related and personal services (1413/7001),
Services for amusement and recreation (1119/8242), Social insurance, social welfare and care services (1119/7616),
Cooperative associations, N.E.C (2366/7925), Miscellaneous services (128/249).
\item {\bf Community rank:} 3, {\bf size:} 59906\\
{\bf Prefecture over-expressions:} Saitama (4665/46418), Chiba (3833/39828), Tokyo (21293/186179), Kanagawa (7438/62781), Osaka (6139/87226).\\
{\bf Sector over-expressions:} Construction work, general including public and private construction work (9653/163082),
Construction work by specialist contractor, except equipment installation work (15242/125183), Equipment installation work (8306/101430),
Production, transmission and distribution of electricity (142/1306), Heat supply (27/80), Banking (40/161), Non-deposit money corporations, 
including lending and credit card business (109/755), Financial products transaction dealers and futures commodity transaction dealers (240/1263),
Financial auxiliaries (69/262), Real estate agencies (3587/19183), Real estate lessors and managers (3570/25301),
Goods rental and leasing (567/6910), Professional services, N.E.C.(1127/19393), Technical services, N.E.C.(2609/26216),
Miscellaneous business services (1887/19963), Foreign governments and international agencies in Japan (7/20).
\item {\bf Community rank:} 4, {\bf size:} 47849\\
{\bf Prefecture over-expressions:} Tokyo (9248/186179), Ishikawa (659/12307), Fukui (1122/11599), Gifu (1127/20230),
Aichi (2929/70128), Kyoto (2407/24369), Osaka (5657/87226), Hyogo (2010/41282), Nara (535/9516), Wakayama (518/10004),
Okayama (1011/20023), Ehime (721/16097).\\
{\bf Sector over-expressions:} Manufacture of textile products (8893/12402), Manufacture of rubber products (222/2060),
Manufacture of leather tanning, leather products and fur skins (1033/1520), Miscellaneous manufacturing industries (958/11270),
Warehousing(112/1997), Wholesale trade, general merchandise ( 536/4891), Wholesale trade (textile and apparel) (7241/10562),
Miscellaneous wholesale trade (1910/34637), Retail trade (woven fabrics, apparel, apparel accessories and notions)(13082/16255),
Miscellaneous retail trade (3107/63111), Nonstore retailers (114/1203), Miscellaneous living-related and personal services (866/7001),
Services for amusement and recreation (691/8242).
\item {\bf Community rank:} 5, {\bf size:} 44349\\
{\bf Prefecture over-expressions:} Hokkaido (2778/58982), Iwate (774/12419), Tochigi (1258/20653), Tokyo(9954/186179),
 Kyoto(1159/24369), Hiroshima (1673/30651), Tokushima (565/9172),  Kochi (394/8078), Fukuoka (2091/45359), Kumamoto (1174/16358),
 Miyazaki (794/12013).\\
 {\bf Sector over-expressions:} Manufacture of chemical and allied product (739/5946), Manufacture of business oriented machinery (575/4194),
 Wholesale trade (machinery and equipment) (1916/40672), Miscellaneous wholesale trade (1735/34637),  Miscellaneous retail trade (6574/63111),
 Scientific and development research institutes (117/547), Medical and other health services (22622/26123), Public health and hygiene (139/387),
 Social insurance, social welfare and care services (1414/7616).
 \item {\bf Community rank:} 6, {\bf size:} 43397\\
 {\bf Prefecture over-expressions:} Aomori (1028/14681), Miyagi (1514/22888), Akita (759/11705),  Yamagata(726/13168), Ibaraki (1211/26412),
 Tochigi (1011/20653), Gunma (1051/19893), Saitama (1819/46418), Chiba (1716/39828), Kanagawa (2380/62781), Toyama (773/13423), Ishikawa (673/12307),
 Fukui (539/11599), Yamanashi (519/9981), Gifu (857/20230), Shizuoka (1514/35806), Mie (741/17747), Okayama (963/20023), Hiroshima (1209/30651),
 Saga (358/7873), Kumamoto (767/16358), Oita (671/12738), Miyazaki (535/12013).\\
 {\bf Sector over-expressions:} Manufacture of transport equipment (369/6623), Road passenger transport (1199/4492),  Wholesale trade
 (machinery and equipment) (3031/40672), Retail trade (machinery and equipment) (15483/41319), Insurance institutions,
 including insurance agents, brokers and services (4805/6234),  Goods rental and leasing (407/6910), School education (312/3416),
 Automobile maintenance services (8607/17703).
 \item {\bf Community rank:} 7, {\bf size:} 43018\\
 {\bf Prefecture over-expressions:} Hokkaido (37867/58982).\\
 {\bf Sector over-expressions:} Agriculture (1214/10747), Forestry (156/934), Fisheries, except aquaculture (62/724), 
 Mining and quarrying of stone (142/1479), Construction work, general including public and private construction work (7077/163082),
Construction work by specialist contractor, except equipment installation work (6095/125183), Equipment installation work (3960/101430), 
Manufacture of lumber and wood products, except furniture (273/5132), Road passenger transport (222/4492), Road freight transport (1554/29678),
Wholesale trade  (building materials, minerals and metals, etc) (1339/34380), Retail trade (food and beverage) (1203/23948), 
Financial institutions for cooperative organizations (41/511), Real estate lessors and managers (1001/25301), 
Goods rental and leasing (336/6910), Technical services, N.E.C. (1266/26216), Accommodations (267/6062),
Eating and drinking places (683/15725), Social insurance, social welfare and care services (328/7616), 
Cooperative associations, N.E.C (520/7925), Automobile maintenance services (1077/17703), Local government services (223/2112).
\item {\bf Community rank:} 8, {\bf size:} 38819\\
 {\bf Prefecture over-expressions:} Tokyo (18716/186179).\\
 {\bf Sector over-expressions:} Printing and allied industries (1138/10996), Manufacture of information and communication electronics equipment (99/2101),
 Miscellaneous manufacturing industries (1312/11270), Communications (75/865), Broadcasting (358/742), Information services (3096/23741),
 Services incidental to internet (427/1236), Video picture information, sound information, character information production and distribution (4361/7665),
 Wholesale trade, general merchandise (246/4891), Miscellaneous wholesale trade (2373/34637), Miscellaneous retail trade (4586/63111),
 Nonstore retailers (256/1203), Financial products transaction dealers and futures commodity transaction dealers (72/1263), 
 Goods rental and leasing (483/6910), Professional services, N.E.C. (2821/19393), Advertising (2572/6306),  Services for amusement and recreation(1225/8242),
 School education (529/3416), Miscellaneous education, learning support (718/2087), Employment and worker dispatching services (558/4683),
 Miscellaneous business services(1701/19963), Political, business and cultural organizations (411/5288).
\item {\bf Community rank:} 9, {\bf size:} 33654\\
 {\bf Prefecture over-expressions:} Tokyo (13884/186179), Kanagawa (2018/62781), Osaka (2667/87226).\\
 {\bf Sector over-expressions:}  Equipment installation work (4096/101430), Manufacture of information and communication electronics equipment (280/2101),
 Production, transmission and distribution of electricity (61/1306), Communications (494/865), Broadcasting (70/742), Information services (11114/23741),
 Services incidental to internet (380/1236), Video picture information, sound information, character information production and distribution (278/7665),
 Wholesale trade (machinery and equipment) (1801/40672), Retail trade (machinery and equipment) (1838/41319), Nonstore retailers(65/1203),
 Banking (17/161), Financial institutions for cooperative organizations (35/511), Non-deposit money corporations, including lending and credit card business (65/755),
 Financial products transaction dealers and futures commodity transaction dealers (272/1263), Financial auxiliaries (49/262),
 Goods rental and leasing (302/6910), Professional services, N.E.C. (2350/19393), Advertising (230/6306), Technical services, N.E.C. (852/26216),
 Miscellaneous education, learning support (128/2087), Employment and worker dispatching services (676/4683), Miscellaneous business services (1154/19963), 
 Political, business and cultural organizations (293/5288).
 
 \item {\bf Community rank:} 10, {\bf size:} 33563\\
 {\bf Prefecture over-expressions:} Gifu (8843/20230), Aichi (19937/70128), Mie (752/17747) \\
 {\bf Sector over-expressions:} Mining and quarrying of stone (88/1479), Construction work, general including public and private construction work (9573/163082),
 Construction work by specialist contractor, except equipment installation work (7621/125183), Equipment installation work (4717/101430),
 Manufacture of lumber and wood products, except furniture (211/5132), Manufacture of ceramic, stone and clay products (354/8051), 
 Wholesale trade  (building materials, minerals and metals, etc) (1168/34380), Real estate agencies (859/19183), Waste disposal business (318/9078),
 Local government services (102/2112).
\end{itemize}
%% References


\bibliographystyle{apsrev4-1}
%\bibliography{com}
   
%\begin{thebibliography}{37}%
%\end{thebibliography}%
\end {document}